\documentclass[plain]{llncs}

\usepackage[T1]{fontenc}
\usepackage{lmodern}

\usepackage[utf8]{inputenc} 
\usepackage[T1]{fontenc}    
\usepackage{xurl}           
\usepackage{hyperref}       
\usepackage{url}            
\usepackage{booktabs}       
\usepackage{amsfonts}       
\usepackage{nicefrac}       
\usepackage{microtype}      
\usepackage[english]{babel}
\usepackage{algpseudocode}
\usepackage{algorithm}
\usepackage{float}
\usepackage{graphicx}
\usepackage[table,xcdraw]{xcolor}
\usepackage{soul}
\sethlcolor{yellow}
\usepackage{multirow}
\usepackage{mathtools}
\usepackage{subcaption}
\usepackage[export]{adjustbox}
\usepackage{cite}
\usepackage{doi}

\usepackage{academicons}
\newcommand{\orcid}[1]{\href{https://orcid.org/#1}{\textsuperscript{\textcolor[HTML]{A6CE39}{\aiOrcid}}}}
\usepackage{tikz}
\usetikzlibrary{svg.path}
\definecolor{orcidlogocol}{HTML}{A6CE39}
\tikzset{
  orcidlogo/.pic={
    \fill[orcidlogocol] svg{M256,128c0,70.7-57.3,128-128,128C57.3,256,0,198.7,0,128C0,57.3,57.3,0,128,0C198.7,0,256,57.3,256,128z};
    \fill[white] svg{M86.3,186.2H70.9V79.1h15.4v48.4V186.2z}
                 svg{M108.9,79.1h41.6c39.6,0,57,28.3,57,53.6c0,27.5-21.5,53.6-56.8,53.6h-41.8V79.1z M124.3,172.4h24.5c34.9,0,42.9-26.5,42.9-39.7c0-21.5-13.7-39.7-43.7-39.7h-23.7V172.4z}
                 svg{M88.7,56.8c0,5.5-4.5,10.1-10.1,10.1c-5.6,0-10.1-4.6-10.1-10.1c0-5.6,4.5-10.1,10.1-10.1C84.2,46.7,88.7,51.3,88.7,56.8z};
  }
}
\newcommand{\OrigHeightRecip}{0.00390625}
\newlength{\curXheight}
\DeclareRobustCommand\orcid[1]{%
\texorpdfstring{%
\setlength{\curXheight}{\fontcharht\font`X}%
\href{https://orcid.org/#1}{\XeTeXLinkBox{\mbox{%
\begin{tikzpicture}[yscale=-\OrigHeightRecip*\curXheight,
xscale=\OrigHeightRecip*\curXheight,transform shape]
\pic{orcidlogo};
\end{tikzpicture}%
}}}}{}}

\hypersetup{
  colorlinks   = true, 
  urlcolor     = blue, 
  linkcolor    = black, 
  citecolor   = black, 
  breaklinks  = true  
}

\usepackage[nonumberlist,acronyms,nogroupskip]{glossaries}

\newcommand{\comment}[1][]{\textcolor{blue}}

\newcommand{\todo}[1][]{\textcolor{red}}
    
\usepackage[utf8]{inputenc}
\usepackage{pgfplots}
\DeclareUnicodeCharacter{2212}{−}
\usepgfplotslibrary{groupplots,dateplot}
\usetikzlibrary{patterns,shapes.arrows}
\pgfplotsset{compat=newest}

\renewcommand{\paragraph}[1]{{\smallskip\noindent {\bf{#1}}~}}

\newacronym{CE}{CE}{cross-entropy}
\newacronym{DL}{DL}{deep learning}
\newacronym{DNN}{DNN}{deep neural network}
\newacronym{HE}{HE}{homomorphic encryption}
\newacronym{CXR}{CXR}{chest X-Ray}
\newacronym{KD}{KD}{knowledge distillation}

\begin{document}


\title{A methodology for training homomorphic encryption friendly neural networks}

\author{%
Moran Baruch\inst{1,2}\orcid{0000-0003-0615-6164} \and 
Nir Drucker\inst{1}\orcid{0000-0002-7273-4797} \and
Lev Greenberg\inst{1}\orcid{0000-0002-1981-9775} \and
Guy Moshkowich\inst{1}\orcid{0000-0003-1856-8430}}

\institute{IBM Research - Haifa \and
Bar Ilan University}

\authorrunning{}%


\maketitle

\thispagestyle{plain}
\pagestyle{plain}

\begin{abstract}
Privacy-preserving \gls{DNN} inference is a necessity in different regulated industries such as healthcare, finance, and retail.  Recently, \gls{HE} has been used as a method to enable analytics while addressing privacy concerns. \gls{HE} enables secure predictions over encrypted data. However, there are several challenges related to the use of \gls{HE}, including \gls{DNN} size limitations and the lack of support for some operation types. Most notably, the commonly used ReLU activation is not supported under some \gls{HE} schemes.

We propose a structured methodology to replace ReLU with a quadratic polynomial activation. To address the accuracy degradation issue, we use a pre-trained model that trains another HE-friendly model, using techniques such as 'trainable activation' functions and \acrlong{KD}. We demonstrate our methodology on the AlexNet architecture, using the \acrlong{CXR} and CT datasets for COVID-19 detection.
Experiments using our approach reduced the gap between the $F_1$ score and accuracy of the models trained with ReLU and the HE-friendly model to within a mere $0.32 - 5.3$ percent degradation. We also demonstrate our methodology using the SqueezeNet architecture, for which we observed $7\%$ accuracy and $F_1$ improvements over training similar networks with other HE-friendly training methods.
\end{abstract}

\section{Introduction}

\begin{figure}[ht!]
\centering
\includegraphics[width=0.7\linewidth]{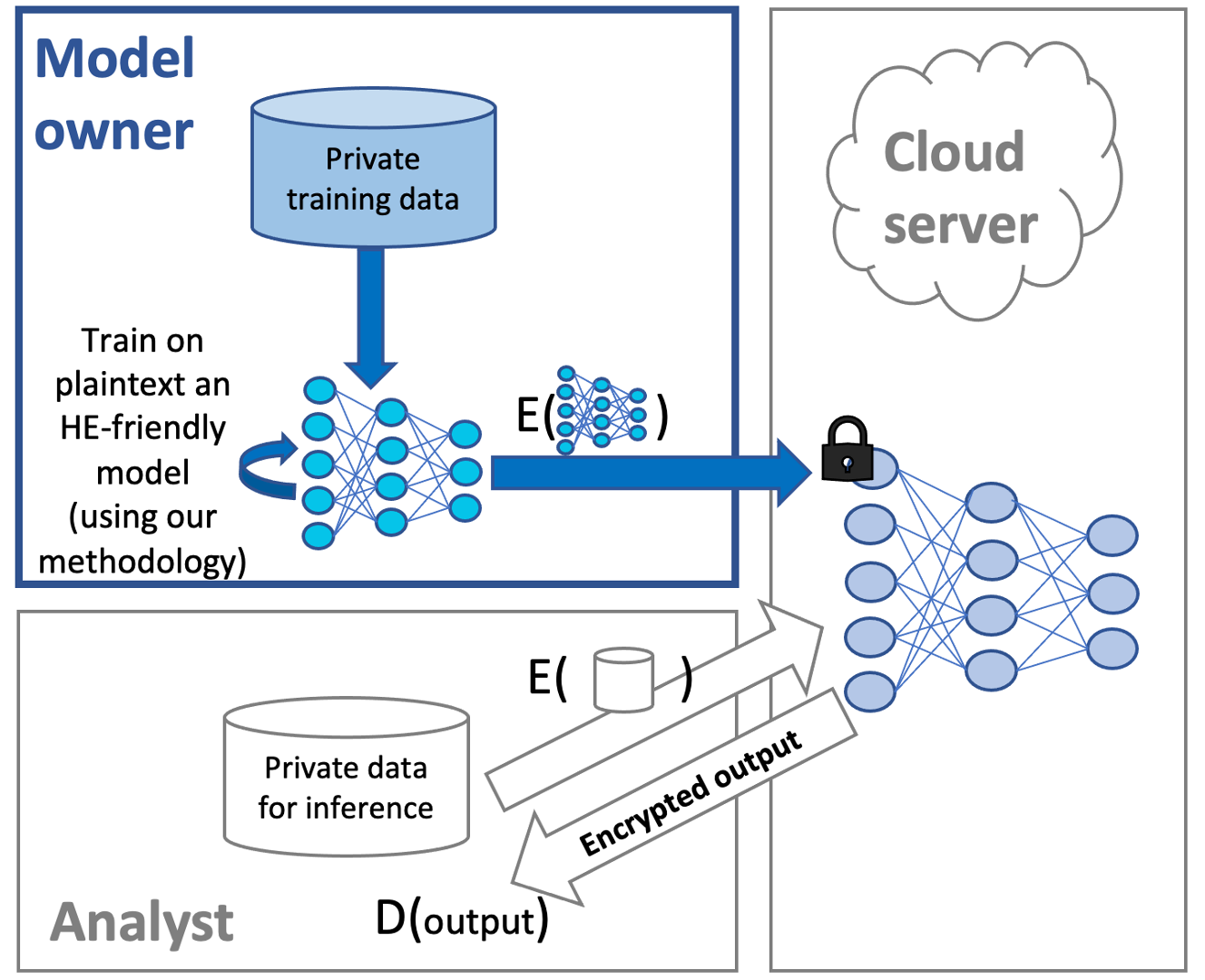}
\caption{This paper focuses on the model-owner training task. A typical flow for running \gls{DNN} over \gls{HE} spans over three entities: a model owner, a cloud server, and an analyst. The model owner \textbf{trains an unencrypted \gls{HE}-friendly \gls{DNN} model}, encrypts it, and uploads it to the cloud. Next, the analyst encrypts some private samples and also uploads them to the cloud. Finally, the cloud processes the encrypted data using the encrypted model and returns the results to the analyst for decryption.}
    \label{fig:he_intro}
\end{figure}

The ability to run \acrfull{DNN} inference on untrusted cloud environments is becoming critical for many industries such as healthcare, finance, and retail. Doing so while adhering to privacy regulations such as HIPAA~\cite{HIPAA} and GDPR~\cite{GDPR} is not trivial. For example, consider a hospital that wishes to analyze and classify medical images  (e.g., \cite{wang2020covid, gunraj2020covidnet}) on the cloud. Regulations may force the hospital to encrypt these images before uploading them to the cloud; this would normally require that the data  first be decrypted before any analytical evaluation can be done.

Homomorphic encryption (HE), which allows computation over encrypted data, is one of the recent promising approaches to help maintain the confidentiality of private data in untrusted environments. At its core, an \gls{HE} scheme provides three capabilities: encryption ($Enc$), evaluation ($Eval$), and decryption ($Dec$). The data owner, say the hospital in our example, can encrypt a message $m$ by invoking $c=Enc(m)$ and then upload the ciphertext $c$ to the cloud, together with some function $f$ that it wishes to evaluate on $m$. Subsequently, the cloud evaluates $c'=Eval(f, c)$ without learning anything about $m$ or the value that $c'$ encrypts. The function returns the encrypted results to the data-owner, who can decrypt it using $m'= f(m)=Dec(c')$ and get the desired results. For further information on \gls{HE}, see \cite{gentry2010computing}. 

\gls{HE} for \gls{DNN} inference is an active research topic   \cite{nandakumar2019towards,gilad2016cryptonets,hesamifard2017cryptodl} focused on using a trained \gls{DNN} model to classify encrypted data. Figure \ref{fig:he_intro} illustrates the overall process and highlights the training phase; this training is done by the model owner on unencrypted data and is the focus of this paper. In practice, the training task is not trivial due to possible limitations of the \gls{HE} scheme.
We describe two principal challenges of \gls{HE} inference.

\paragraph{Multiplication depth.}
Multiplication depth is defined as the longest chain of sequential multiplication operations in the \gls{HE} evaluated function. Some \gls{HE} schemes only allow for a certain number of consecutive multiplication operations. To tackle this challenge, such schemes use a \textit{bootstrapping} operation \cite{gentry2009fully} that allows further computation. 
Because bootstrapping is expensive in terms of run-time, reducing the multiplication depth allows us to reduce or avoid bootstrapping, while speeding up the entire computation. 

\paragraph{Non-polynomial operations.} Some modern \gls{HE} schemes support only basic arithmetic operations of addition and multiplication e.g., CKKS \cite{ckks2017} and BGV \cite{bgv2014} schemes. Consequently, only  \gls{DNN} components that can be represented as a composition of these arithmetic operations can be computed directly in \gls{HE}.

One way to overcome this limitation is by using a polynomial approximation to approximate the operation.
For example, the ReLU activation function defined as $\text{ReLU}(x)=\max(0,x)$ is approximated by a polynomial in \cite{lee2021privacy_approxhigh, takabi2019privacy,secureml2017,hesamifard2017cryptodl}.
A second option is to replace the operation with a similar but different \gls{HE}-friendly operation. For example, this may involve replacing a \textit{max-pooling} operation with the \gls{HE}-friendly operation of \textit{average-pooling}, which in many use cases does not affect the \gls{DNN} performance \cite{gilad2016cryptonets}.
A third option, is to use a client-aided design \cite{lloret2021enabling}, where the hard-to-compute operation is sent to the data-owner who decrypts the data, computes the operation, encrypts the result, and sends it back to the cloud to continue its \gls{HE} computation. We prefer to avoid this method because, in addition to the communication complexity, 
it increases the attack surface and opens the door to theoretical attacks such as those suggested by Akavia et al. \cite{akaviaprivacy} or model-extraction attacks as presented by Li \cite{muse}. We summarize our research question.

\paragraph{Research question.} Can we find a methodology for modifying \gls{DNN} architectures and their training process to produce a \gls{HE}-friendly model with similar prediction accuracy as the original \gls{DNN}?

\subsection{Our Contributions}
We propose a new methodology that combines several techniques for adapting and training \gls{HE}-friendly \glspl{DNN} on the plaintext, to enable homomorphic inference over encrypted data. 
In these \glspl{DNN}, we replace the ReLU activations by customized quadratic polynomial activation functions. 
Our methodology also enables the entire inference process to occur in the cloud environment, without interaction with the data-owner.
We show empirically that the resulting inference accuracy is comparable with the inference accuracy of our \textit{baseline}, the original \gls{DNN} with the ReLU activation function.

Our customized activation functions apply the following techniques:
\begin{itemize}
    \item Low-degree polynomial activation functions with trainable coefficients
    \item Method for gradual replacement of the original activation functions during the training phase
    \item Adaptation of the \gls{KD} technique \cite{hinton2015distilling} to train an \gls{HE}-friendly model from a pre-trained baseline model in its vanilla settings
\end{itemize}
We evaluated the efficiency of our method on two different model architectures, AlexNet \cite{krizhevsky2012alexnet} and SqueezeNet \cite{squeezenet2016}, for the task of COVID-19 detection over CT and \gls{CXR} images of size $224 \times 224 \times 3$. We chose these datasets for their relevance to the current pandemic, as it may enable hospitals to evaluate COVID-19 cases on the cloud, and analyze them globally. In addition, we prefer these datasets over more standard datasets such as MNIST or CIFAR-10, which have much smaller image sizes: $28 \times 28 \times 1$ and $32 \times 32 \times 3$, respectively; as a result, their \gls{DNN} models are also much smaller. For completeness, we also evaluated our methodology on CIFAR-10 images that were resized to $224 \times 224 \times 3$, and showed that our methodology outperforms previous works, even when using the original AlexNet. 

Our results for AlexNet demonstrated a minimal degradation of up to $5.3\%$ in the $F_1$ score, compared to the original baseline models. For both architectures, we improved the $F_1$ score by $4\%-10\%$ compared with HE-friendly networks that we trained using state-of-the-art methods. Note that we chose to demonstrate our methodology on AlexNet and SqueezeNet as these are, to the best of our knowledge, the deepest \gls{HE}-friendly architectures that were demonstrated to run encrypted within a reasonable amount of time over large images. Other architectures such as VGG-16, MobileNet, and ResNet-20 were either demonstrated for non-HE-friendly, client-aided solutions \cite{ngraph} or over datasets with smaller images such as CIFAR-10 with images of size $32\times 32 \times 3$ as in \cite{duran}.
Our trained models are available online\footnote{https://ibm.ent.box.com/folder/161803670185?v=fhe-friendly-models}.


\paragraph{Organization.}
The paper is organized as follows. Section \ref{sec:related_work} surveys the relevant literature. Section \ref{sec:our_methods} presents our methodology and the techniques we used. We present our experiments in Section \ref{sec:experiments} and conclude in Section \ref{sec:conc}.

\section{Related Work} \label{sec:related_work}

The ReLU function uses the non-polynomial \textit{max} operation, which is not supported by some \gls{HE} schemes such as CKKS. These schemes can only address this limitation using methods such as lookup tables and polynomial approximations. 

Using lookup tables to approximate ReLU was introduced in \cite{piazza1992artificial, meher2010optimized}, and was used to homomorphically train \glspl{DNN} in \cite{lou2019glyph,nandakumar2019towards}. One disadvantage of this approach is the low resolution of the lookup table, which is limited by the number of lookup table entries. This number is significantly lower than the number of values possible in a single or double floating-point number. In addition, this technique is not available for all \gls{HE} schemes, such as CKKS.

The second approach involves techniques to replace the ReLU activation function with a polynomial approximation function. This can be done using an analytical method to approximate the polynomial or machine learning to train the polynomial coefficients. 
The work of Cheon et al. \cite{cheon2019numerical} describes a method for approximating the generic \textit{max} function. However, using this method often leads to a high degree polynomial approximation, which increases the multiplication depth and the accumulated noise. In addition, this approximation is applicable only when the input operands are limited to a specific range. Unlike the above methods that can approximate generic functions, we are interested in ReLU. ReLU is a special case of the \textit{max} function, where one of the \textit{max} input operands is fixed to zero. Hence, it is possible to use other approximation methods that yield polynomials with even lower degrees and better performance, while improving the overall efficiency. 

The square function ($square(x) = x^2$) \cite{gilad2016cryptonets} is a well-known low-degree polynomial replacement for the ReLU function. However, when the number of layers in the model grows, the accuracy of the model degrades significantly. 
To mitigate this degradation, several works \cite{lee2021privacy_approxhigh,takabi2019privacy, secureml2017, hesamifard2017cryptodl} suggested using a higher-degree polynomial, which again leads to high multiplication depth. For example, Lee et al. \cite{lee2021privacy_approxhigh} used polynomial approximation with degrees $15$ and $27$. However, these polynomials had to use bootstrapping twice for each activation function. The excessive use of bootstrapping caused them to report the results only for a 98-bit secure solution. In contrast, our method enabled the authors of \cite{helayers} to run the AlexNet model on large images using 128-bit security and without any bootstrap operations.

Another mitigation suggested by Wu et al.\cite{takabi2019privacy} approximated ReLU using the quadratic polynomial  $0.00047x^2 + 0.5x$ instead of a simple square function. To evaluate the performance of their methods, the authors use a lighter variant of AlexNet \cite{krizhevsky2012alexnet} with images of size $32 \times 32 \times 3$. We tested their approach on the original AlexNet architecture with larger images of size $224 \times 224 \times 3$. As reported in Section \ref{sec:experiments}, this approximation suffers from a degradation in accuracy of up to $35\%$.

The studies above searched for a ReLU replacement that would serve as a good polynomial approximation. They then replace all the ReLU occurrences in a model with this approximation. In contrast, we consider a fine-tuning approach, in which we use a different activation per layer, without necessarily approximating the ReLU activation function. To this end, we used a \gls{DNN} to train the coefficients of the different polynomials. We call this technique 'trainable activation' \footnote{Other names suggested in previous works are ``parametric activation'' \cite{wu2018ppolynets} and  ``adaptive polynomial activation'' \cite{scardapane2017learning,piazza1992artificial,zhang2002neuron}}. 

A similar approach was suggested in other works \cite{piazza1992artificial, zhang2002neuron, scardapane2017learning}, in which the authors trained a polynomial per neuron in small networks of several dozens of neurons. Clearly, this approach is not feasible in modern networks, where the number of neurons is in the order of millions and the number of parameters requiring optimization is huge. 

Instead of training a polynomial per neuron, Wu et al. \cite{wu2018ppolynets} suggested training a polynomial per layer, for all channels together. They evaluated their activation functions on 3 models, where the largest model has 4 convolutional layers, 2 average-pooling layers, 2 fully-connected layers, and 3 polynomial activations, accompanied by a batch normalization layer. The experiments were applied on the MNIST dataset, which consists of images of size $28 \times 28 \times 1$. We scaled up this approach by using a larger model (see Appendix \ref{appendix:a}), over larger images, and extended their methodology by combining additional techniques to help the model converge better. Our evaluation shows that we were able to narrow down the accuracy gap between the HE-friendly AlexNet and SqueezeNet, trained using the approach of \cite{wu2018ppolynets} with our baseline. Our experiments show that we improved the performance of the model by up to $12.5\%$ compared to their approach.

\section{Methodology} \label{sec:our_methods}

\begin{figure}[ht!]
\centering
\includegraphics[width=0.7\linewidth]{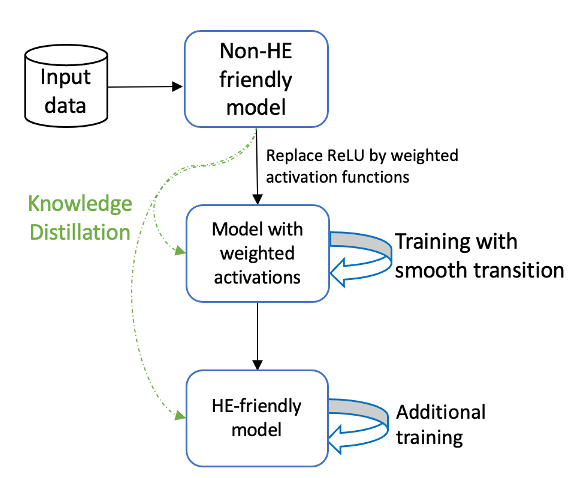}
\caption{Training methodology for HE-friendly \glspl{DNN}.}
    \label{fig:methodology}
 \end{figure}

Our goal was to replace the ReLU function with a polynomial activation function. This section describes the training methodology we used to achieve comparable performance with the baseline model, as presented in Figure \ref{fig:methodology}.

\subsection {Trainable Polynomial Activation}\label{trainable_polynomial}
Based on our approach, we wanted to design a trainable polynomial that would replace the ReLU activation, without approximating it.
Recent papers suggested approximating ReLU using high-degree polynomials to achieve a good approximation. 
However, for a polynomial of degree $n$, this requires an order of $log_2(n)$ multiplications \cite{SCHONHAGE19751}, which significantly increases the computation depth as the number of multiplications grows.

Therefore, we suggest using a trainable polynomial activation of a $2^{nd}$ degree polynomial without the constant term. We used the form $ax^{2} + bx$, where $a$ and $b$ are trainable coefficients, which we trained  individually \textbf{per layer}. A similar approach was presented by Wu et al. \cite{wu2018ppolynets}, where each such activation layer only increments the multiplication depth by 1.

\subsection{Smooth-Transition} \label{section:ST}
Applying such a significant architectural change to a complex model, without first adapting the model weights, can lead to a steep drop in accuracy. Hence, we designed a new approach that we call \textit{smooth-transition}. We start by training a model that includes ReLU activation layers for $e_0$ epochs. Over the next $d$ epochs, we smoothly transition from the ReLU functions to the polynomial activation functions $poly\_act()$. Then, we continue to train the model on the transitioned \gls{HE}-friendly architecture. To model this, we use the ratio parameter $\lambda_e$ per epoch $e$.
\begin{equation*}
    \lambda_e = \begin{dcases}
    0,              &  (e - e_0) \leq 0\\
    \frac{e - e_0}{d} &  0 < (e -e_0) < d\\
    1 & Otherwise
\end{dcases}
\end{equation*}
and set the \textbf{weighted activation function} at epoch $e$ as $$weighted\_act_{\lambda_e}(x) := (1-\lambda_e) \cdot ReLU(x)+ \lambda_e \cdot poly\_act(x).$$ 
To help the network converge, we initiated the quadratic function $poly\_act(x)=ax^2+bx$ as a linear function that is somewhat similar to ReLU, by setting $a=0$ and $b=1$. We stress that the weight $\lambda_e$ is not trained during the transition phase, instead it is predefined according to the smooth-transition policy.

\begin{remark}
We tried replacing the ReLU activation functions with quadratic polynomials layer-by-layer, instead of replacing all layers in parallel. However, it did not provide any significant advantage, and in some cases even showed performance degradation.
\end{remark}


\subsection{Knowledge Distillation} \label{subsec:distil}
Using polynomial activations instead of ReLU activations is less suitable for the classification task. To strengthen the model, we adopted the well known \gls{KD} \cite{hinton2015distilling} technique.

\gls{KD} enables a knowledge transfer from a stronger  pre-trained 'teacher' model to a weaker 'student' model. In practice, the student model is usually smaller than the teacher \cite{mirzadeh2020improved}. In our case, replacing  ReLU by a polynomial activation weakens the HE-friendly model; therefore, the original model is used as a 'teacher' model to assist in training the 'student' HE-friendly model.

We used the response-based \gls{KD} approach, one of the simplest \gls{KD} methods \cite{gou2021knowledge}.
Here, an additional term is added to the loss function to measure discrepancies between the predictions of the teacher and student models. 
We also employed the \textit{soft target} technique \cite{hinton2015distilling}, in which \textit{soft targets} are used instead of the original predictions of the teacher and student models:
\begin{equation*} 
Q^\tau[i] = \frac {\exp{(z_i/\tau)}}{ \sum_j \exp{(z_j/\tau)}}
\end{equation*}
where $Q^\tau[i]$ is the \textit{soft target} version of the prediction for the class $i$, $z_i$ are the original prediction logits, and $\tau$ is the \textit{temperature} \cite{hinton2015distilling}.
With $\tau=1$, the above formula becomes the standard ``softmax'' output: using a higher temperature value ($\tau>1$) produces a more uniform distribution of the probabilities over the classes.
The resulting loss function becomes \cite{liexploring,KD2021Pytorch}:
\begin{equation*}
 L_{KD} = \alpha \tau^{2} \cdot CE (Q^\tau_s, Q^\tau_t) + (1-\alpha) \cdot CE(Q^1_{s}, y_{true})
\end{equation*}
where $Q^\tau_s$ and $Q^\tau_t$ are vectors of the soft target predictions of the student and teacher models with the same temperature $\tau>1$, $Q_{s}^1$ is the ``softmax'' student prediction, $y_{true}$ is the original labels, $CE$ is the cross-entropy loss function, and $\alpha$ is the hyperparameter controlling the relative weight of the additional \gls{KD} loss term.
 
\section{Experiments} \label{sec:experiments}
 
\subsection{Datasets} 

Our experiments use two datasets: COVIDx and COVIDx CT-2A.
\begin{itemize}
\item \textbf{COVIDx \cite{wang2020covid}.}
This is a dataset of \gls{CXR} images labeled as: \textit{Normal}, \textit{Pneumonia}, or \textit{COVID-19}. It is an open access benchmark dataset comprising ${\sim}20,000$ \gls{CXR} images, with the largest number of publicly available COVID-19 positive cases. This dataset collects its data from 6 chest X-Ray datasets \cite{cohen2020covid, covid19chest,tsai2021rsna,9144185,clark2013cancer,rahman2021exploring} and combines them into a big dataset that is updated over time with more COVID-19 positive \gls{CXR} images. The number of images we used per class is depicted in Table \ref{datasets_description}.
  When creating this dataset, we verified that there are no patients overlapping between the train, test, and validation subsets.
We applied an augmentation process to the data, similar to Wang et al. \cite{wang2020covid}.

\item \textbf{COVIDx CT-2A \cite{gunraj2020covidnet}.}
   This dataset contains 194,922 chest CT slices from 3,745 patients, with the same classes as in the previous dataset. We used a random balanced subset of the original dataset, as depicted in Table \ref{datasets_description}.
Each image was augmented as follows: resize to $224 \times 224 \times 3$, random rotation, horizontal flip, vertical flip, color jitter, and normalize. 
\end{itemize}

\begin{table}[ht!]
\centering
\caption{COVIDx and COVIDx CT-2A data sizes used per class}
\label{datasets_description}
\begin{tabular}{l|l|c|c|c|c|}
\cline{3-6}
                               \multicolumn{2}{l|}{} & \textbf{Normal} & \textbf{Pneumonia} & \textbf{COVID-19} & \textbf{Total} \\ 
\hline
\multicolumn{1}{|l|}{\multirow{3}{*}{\textbf{COVIDx}}}       & Train          & 7966            & 5475               & 4303              & 17744          \\ \cline{2-6} 
\multicolumn{1}{|l|}{}                                       & Validation     & 797             & 534                & 559               & 1871           \\ \cline{2-6} 
\multicolumn{1}{|l|}{}                                       & Test           & 88              & 60                 & 61                & 209            \\ 
\hline
\hline
\multicolumn{1}{|l|}{\multirow{3}{*}{\textbf{COVIDx CT-2A}}} & Train          & 10000           & 10000              & 10000             & 30000          \\ \cline{2-6} 
\multicolumn{1}{|l|}{}                                       & Validation     & 1000            & 1000               & 1000              & 3000           \\ \cline{2-6} 
\multicolumn{1}{|l|}{}                                       & Test           & 100             & 100                & 100               & 300            \\ \hline
\end{tabular}
\end{table}

\subsection{Model}
For evaluation, we used the AlexNet \cite{krizhevsky2012alexnet} and SqueezeNet \cite{squeezenet2016} models. We chose AlexNet because it was the deepest network that was tested for non-interactive HE solutions \cite{helayers}. We also used SqueezeNet, which was designed as a light version of AlexNet with $50\times$ fewer parameters. Although lighter, it is a much deeper network with $40$ layers instead of $21$. The models were originally pretrained on the ImageNet \cite{imagenet} dataset, and then fine-tuned on the COVIDx datasets. Because both original models are not HE-friendly, we describe the steps to transform the original models into HE-friendly models.

\subsubsection{AlexNet}
We implemented an AlexNet model based on PyTorch\footnote{https://pytorch.org}, and added a batch normalization layer after every activation layer. To avoid additional multiplication depth, after the training process ended, we absorbed the coefficients of the batch normalization into the weights of the next layer, as suggested by Ibarrondo and Onen \cite{ibarrondo2018batchnorm}. Appendix A presents the network architecture.

HELayers \cite{helayers} is an AI over HE framework. Following the pre-print version of this paper \cite{baruch2021fighting}, the developers of HELayers tested our methodology for AlexNet. Their results showed a speedup for  time and accuracy over encrypted input when using large networks (in terms of HE) and large image sizes of $224 \times 224 \times 3$.

\subsubsection{SqueezeNet}
The SqueezeNet model \cite{squeezenet2016} aims to achieve AlexNet-level accuracy with $50\times$ fewer parameters. This comes at the cost of significantly increasing the multiplication depth, from $21$ layers to $40$ layers. Unlike our approach for AlexNet, we did not add \textit{batch normalization} for SqueezeNet, as we did not observe significant performance improvement when using it. Our model architecture is the SqueezeNet version 1.0 implemented in PyTorch. 

A lighter HE-Friendly version of SqueezeNet tailored for CIFAR-10 with $23$ layers instead of $40$ over small images of size $32 \times 32 \times 3$ was implemented and evaluated by Dathathri et al. \cite{CHET2019}. The successful implementation of this lighter version increased our motivation to offer a method that can also successfully train the original (larger) SqueezeNet over larger images while achieving acceptable accuracy. 

Our evaluation is focused on AlexNet and SqueezeNet as these are the largest model architectures that were demonstrated to run over HE in a non-interactive mode, i.e., without using client-aided designs. We are not aware of other attempts to use larger networks while also considering large image sizes as in our case.

\subsection{Experimental Results}

\subsubsection{AlexNet}
Table \ref{table:alexnet_results} summarizes our experimental results using different methods on the COVIDx and COVIDx-CT-2A datasets. In every experiment, we measured the accuracy and macro-average of the $F_1$ scores on all of the three classes. We repeated every experiment five times with different seeds and report the average results and the standard deviation. For more details regarding the training setup, see Appendix B.

As can be seen from Table~\ref{table:alexnet_results}, 
the \textbf{trainable activation} improved the results by 14 - 20\% when compared to the quadratic approximated ReLU. The \textbf{smooth-transition (ST)} approach, which gradually changes the activation function over 10 epochs starting from the $3^{rd}$ epoch, further improved the results by 3.4 - 8.6\% when compared to replacing all activations at once. Finally, combining both approaches with \textbf{Knowledge Distillation (KD)}, where the original AlexNet with ReLU was used as a teacher to the new adapted architecture, performed even better with an improvement of 1.3 - 3.6\% (baseline divided by our method). This almost closed the gap with the original reference model, with only 0.32 - 5.3\% degradation.

\begin{table}[t!] 
\centering
\caption{A comparison of our suggested methods and their contributions to previous works, over the AlexNet model. The results are reported for the test data of COVIDx and COVIDx CT-2A images. The baseline network is the original network with max pooling and ReLU. For all columns, higher values are better. We use the term TP for Trainable Polynomials,  ST for Smooth-Transition and KD for Knowledge Distillation.}
\label{table:alexnet_results}
\begin{tabular}{|l|cc|cc|}
\cline{2-5}
  \multicolumn{1}{l|}{}
  & \multicolumn{2}{c|}{\textbf{COVIDx CT-2A}}  
  & \multicolumn{2}{c|}{\textbf{COVIDx}} \\ 
  \hline
    \textbf{Technique}
  & \textbf{Accuracy}
  & {$\mathbf{F_1}$}
  & \textbf{Accuracy}
  & $\mathbf{F_1}$ \\ 
  \hline
    Square \cite{gilad2016cryptonets} & $0.435\pm 0.11$ & $0.429 \pm 0.13$ & $0.378 \pm 0.06$ & $0.372 \pm 0.08$ \\ 
    Approx. ReLU \cite{takabi2019privacy}
  & $0.706 \pm 0.02$ & $0.703 \pm 0.09$ & $0.696 \pm 0.01$ & $0.670 \pm 0.01$ \\ \hline 
  {TP}  \cite{wu2018ppolynets}
  & $0.806 \pm 0.03$ & $0.807 \pm 0.03$ & $0.811 \pm 0.02$ & $0.809 \pm 0.04$ \\ 
    \textbf{Our method TP+ST}
  & $0.837\pm0.01$ & $0.835\pm0.10$ & $0.881\pm0.12$ & $0.878\pm0.14$ \\ 
    \textbf{Our method TP+ST+KD}
  & $\mathbf{0.848\pm0.04}$ & $\mathbf{0.847\pm0.09}$ & $\mathbf{0.913\pm0.10}$ & $\mathbf{0.907\pm0.16}$ \\ 
  \hline
    \multicolumn{1}{|l|}{Baseline}
  & $0.893\pm0.03$ & $0.892\pm0.08$ & $0.916\pm0.01$ & $0.915\pm0.03$ \\ 
    \hline
\end{tabular}
\end{table}

 \begin{table}[t!] 
\centering
\caption{A comparison of our suggested methods and their contributions to previous works over the SqueezeNet architecture. The results are reported on the test data of COVIDx CT-2A images. The baseline network is the original network with max-pooling and ReLU; we also added a reference model with average-pooling and ReLU. For all columns, higher values are better. }
\label{table:squeezenet_results}
\begin{tabular}{|l|cc|} 
\hline
\textbf{Technique}           & \textbf{Accuracy}                               & $\mathbf{F_1}$                                    \\ 
\hline
Square                       & $0.33\pm 0.00$                                  & $0.33 \pm 0.00$                                   \\
Approx. ReLU                 & $0.740 \pm 0.15$                                & $0.728 \pm 0.15$                                  \\ 
\hline
TP                           & $0.754 \pm 0.05$                                & $0.742 \pm 0.06$                                  \\
\textbf{Our method TP+ST}    & $0.806\pm0.23$                                  & $0.800\pm0.05$                                    \\
\textbf{Our method TP+ST+KD} & $\mathbf{0.820\pm0.23}$                         & $\mathbf{0.816\pm0.02}$                           \\ 
\hline
Baseline (ReLU + Avgool)     & $0.825\pm0.23$                                 & $0.826\pm 0.31$  \\
Baseline (ReLU + Maxpool)      & $0.898\pm0.01$                                  & $0.897\pm0.05$                                    \\
\hline
\end{tabular}
\end{table}

\subsubsection{SqueezeNet}
Table \ref{table:squeezenet_results} summarizes the results of using different methods on the COVIDx-CT-2A dataset. The results are reported in a similar format as in Table \ref{table:alexnet_results}. The only difference is that in this table we also added a baseline model, which consists of the original model of SqueezeNet with ReLU but with average-pooling. The reason is that when we replaced the max-pooling layer with an average-pooling layer, both the $F_1$ scores and the accuracy scores went down by around $8\%$. This observation is interesting because it shows that the claim of Gilad-Backrach et al. \cite{gilad2016cryptonets} does not hold for all models. This may also open the door for a new line of research that focuses not only on the activation layers but also on max-pooling layers. 

We compared the accuracy and $F_1$ score of models with average-pooling, where one is trained using our methodology and the other using the standard methodologies. The results showed almost the same performance as with AlexNet, with $0.6-1.2\%$ degradation in the former model. Interestingly, the smooth-transition technique played an important role, and improved the results by $6.9\%-7.8\%$. Interestingly, when we evaluated a model with square activations, the model did not converge at all. 

\paragraph{CIFAR-10 for AlexNet}. For completeness, we also evaluated our methodology on the well-known CIFAR-10 dataset over the AlexNet model. We resized the images to $224 \times 224 \times 3$ to fit the input size required by the model. The baseline network with ReLU reached an accuracy of 0.901, and the accuracy of the HE-friendly model using our full methodology was only slightly lower at 0.872 (and 0.869 without \gls{KD}). Using non-smooth trainable activation functions resulted in a low accuracy of 0.751, and with an approximated ReLU it reached  0.7405. Again, we see that our methodology outperforms previous works, especially when evaluating on the original AlexNet model without modifications, and over large images.

\subsubsection{Varying polynomial activations per layer}
\begin{figure}[t!]
    \centering
    \begin{subfigure}[b]{0.49\textwidth}
        \centering
        \includegraphics[width=0.98\textwidth, frame]{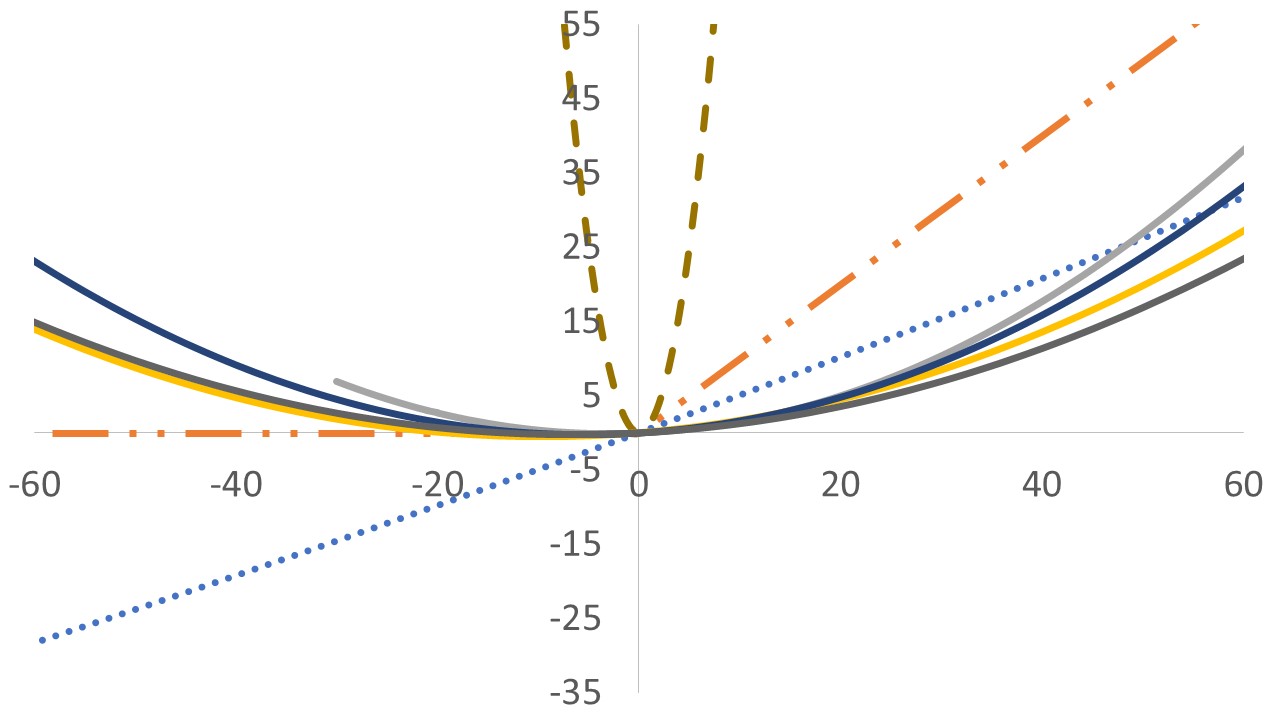}
        \caption{AlexNet $[-60,60]$}
    \end{subfigure}
    \begin{subfigure}[b]{0.49\textwidth}
        \centering
        \includegraphics[width=0.98\textwidth, frame]{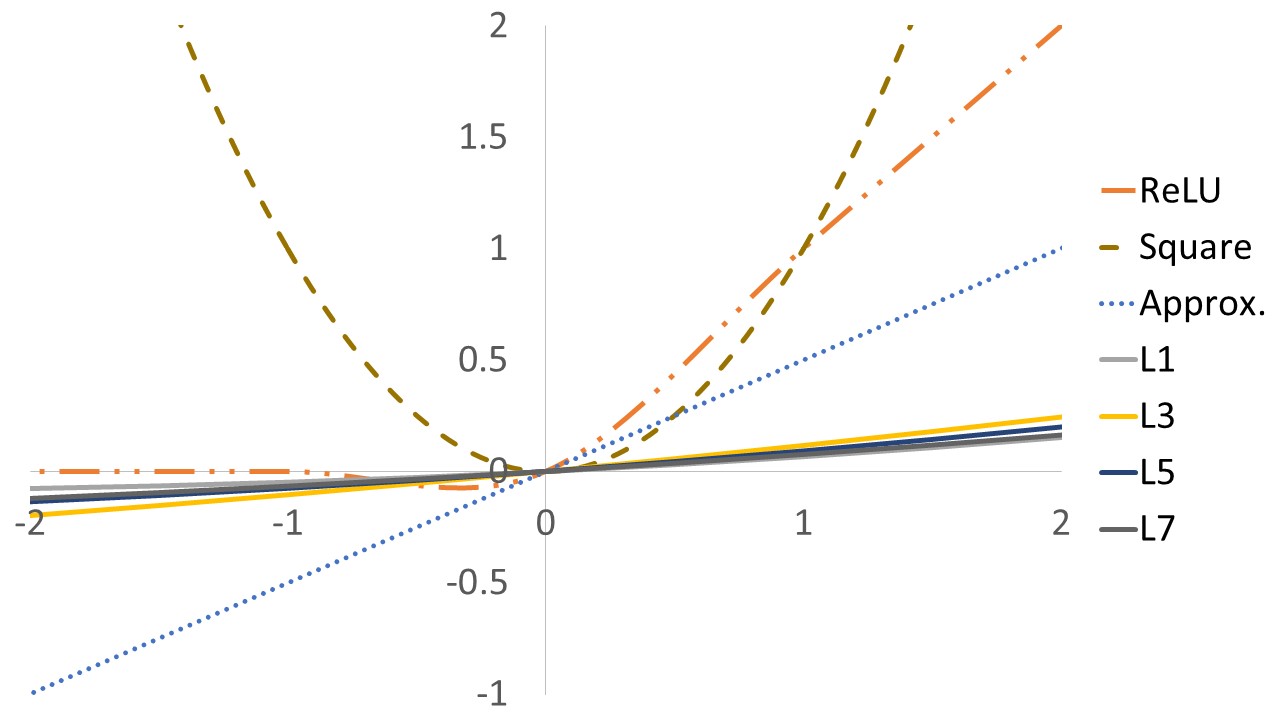}
        \caption{AlexNet $[-2,2]$}
    \end{subfigure}
    
     \qquad
    
     \begin{subfigure}[b]{0.49\textwidth}
         \centering
         \includegraphics[width=0.98\textwidth, frame]{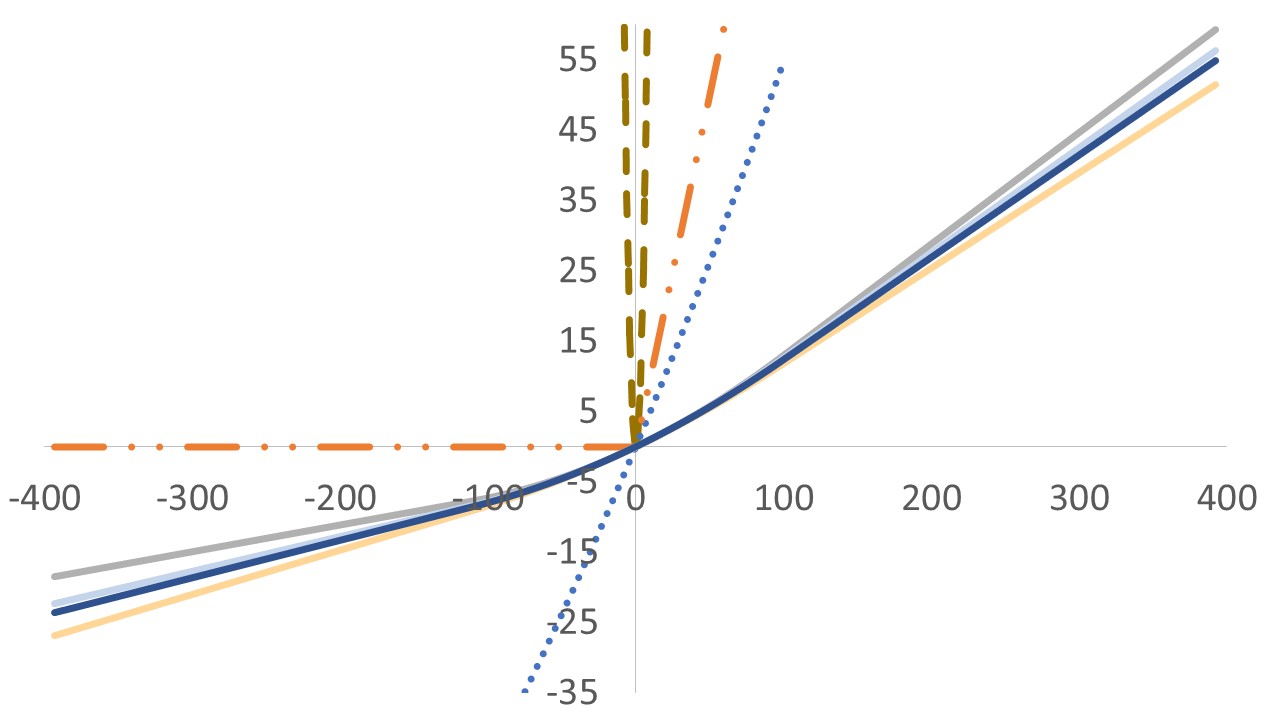}
         \caption{SqueezeNet $[-400,400]$}
     \end{subfigure}
     \begin{subfigure}[b]{0.49\textwidth}
         \centering
         \includegraphics[width=0.98\textwidth, frame]{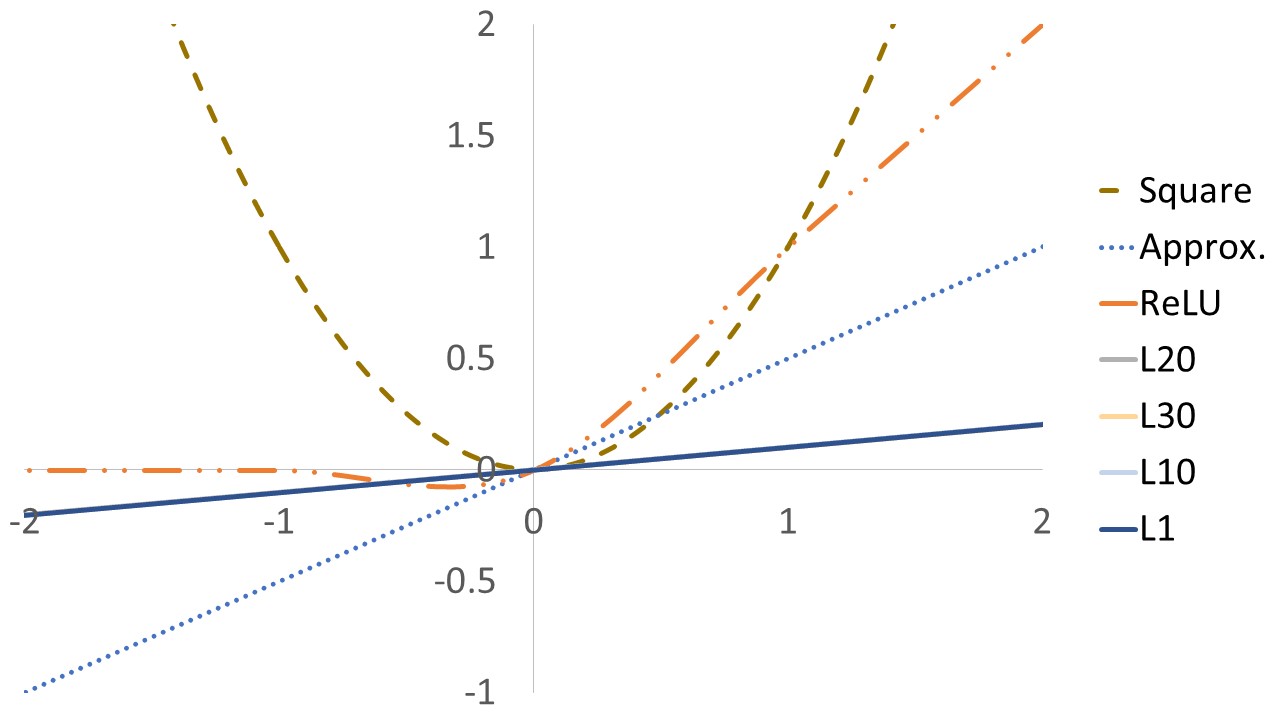}
         \caption{SqueezeNet $[-2,2]$}
     \end{subfigure}
    \caption{Different activation functions: ReLU, Square function, ReLU approximation and the resulted trainable polynomials for several layers of AlexNet (top panels) and SqueezeNet (bottom panels). In panel (d) the curves of L1-L30 coincide.}
    \label{fig:trainable_activations_graphs}
\end{figure}

To better understand the final activation polynomials ($ax^2+bx$), we analyzed the ranges of their coefficients. For AlexNet, we got $a \in [0.003, 0.010]$, with a standard deviation of $32\%$ of the $0.0065$ average value and $b \in [0.057,0.110]$, with a standard deviation of $20\%$ of the $0.0830$ average value. We got similar results for SqueezeNet. Figure \ref{fig:trainable_activations_graphs} presents a sample of the activation function graphs of different layers next to the graphs of the ReLU, square, and the ReLU approximation functions. The large variance between the polynomials may explain the accuracy advantage we see when using several different polynomial activations compared to using only one fixed approximation for all layers. 

In our experiments, the inputs to the AlexNet activation functions were in the range [-90, 90], the average input value for most layers was close to zero with standard deviations in the range $[4, 18]$. In contrast, in SqueezeNet, the input to the activation functions were in $[-100,100]$ for the first 18 activations, and in $[-1506,1356]$ otherwise. Here too, the average input value for most layers was close to zero with standard deviations of less than $20$ for the first 18 layers and around 180 for the other layers. Thus, Figure \ref{fig:trainable_activations_graphs} shows the graphs for small inputs in $[-2,2]$ but also in wide ranges $[-60,60]$ and $[-400,400]$ for AlexNet and SqueezeNet, respectively. 

The graphs show that, unlike ReLU, the square function grows and even explodes for negative inputs, therefore it is less likely to cause a network to converge. In contrast, compared to ReLU, the ReLU approximation outputs lower values for both negative and positive inputs. This allows the network to converge, but the increased weight it gives to negative inputs might be the reason for the observed lower accuracy. Interestingly, our trainable functions happen to be closer to ReLU, at least in the range $[-2,0]$ and starts to deviate from it when extending the input range. It seems that in $[-2,2]$ with small inputs, our trainable functions almost agree on their outputs, but for distant inputs, they uniquely define the characteristics of the layers. In our experiments, we did not observe that the order of layers dictates some order on the trained functions.

\subsubsection{Smooth transition and training robustness}
\begin{figure}[ht!]
\centering
\begin{subfigure}[b]{0.98\linewidth}
        \centering
        \includegraphics[width=\linewidth]{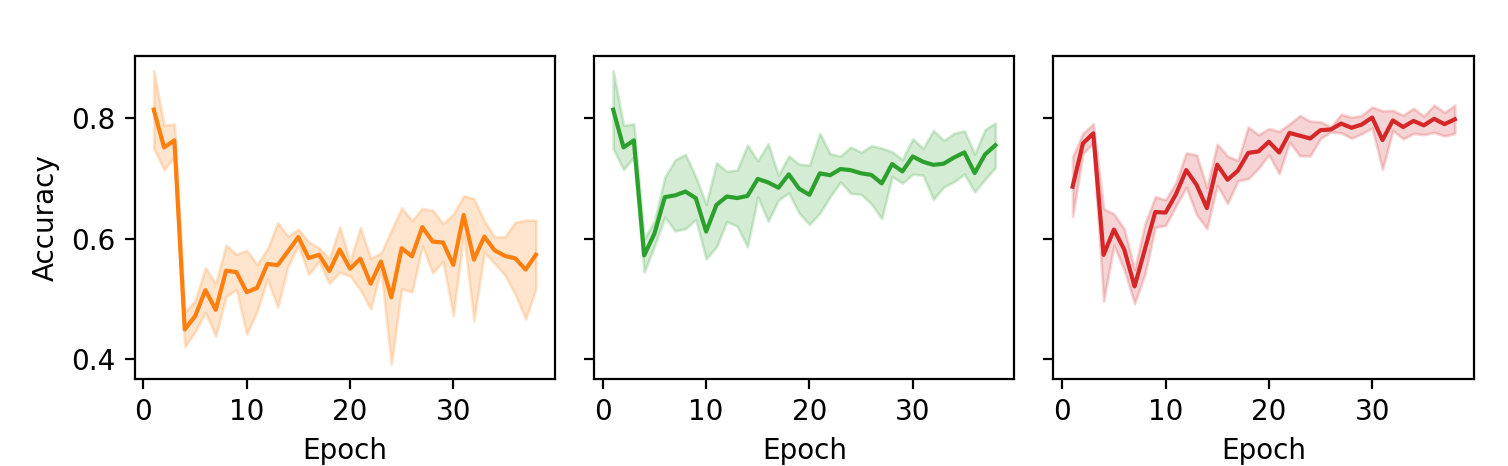}
        \caption{AlexNet}
    \end{subfigure}
    
    \begin{subfigure}[b]{0.98\linewidth}
        \centering
        \includegraphics[width=\linewidth]{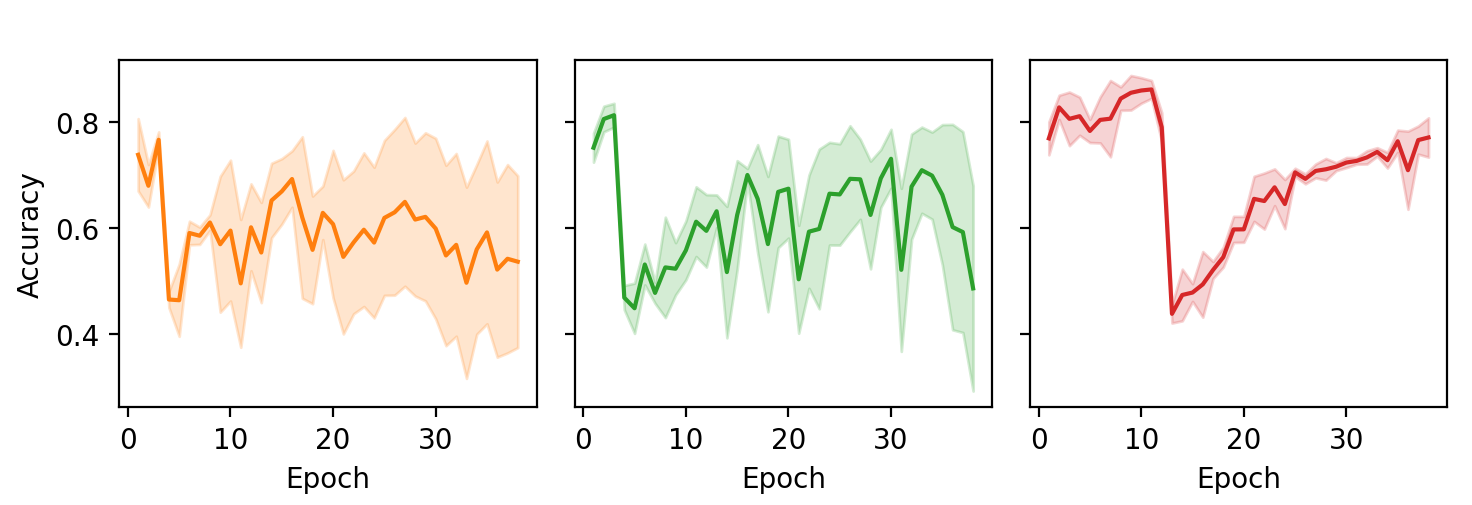}
        \caption{SqueezeNet}
    \end{subfigure}
    \caption{Test accuracy for the AlexNet (top) and SqueezeNet (bottom) architectures trained with the approximated ReLU (left, orange), trainable polynomial without smooth-transition (middle, green), and trainable polynomial with smooth-transition (right, red) -- our method. The graphs show the average and spread of five different runs with different seeds. A smaller spread indicates a more stable training method.}
    \label{fig:robustness_squeezenet}
 \end{figure}

Figure \ref{fig:robustness_squeezenet} compares the robustness of the previous methods for replacing the ReLU activations with our smooth transition approach over five different seeds. The graphs show that using approximated ReLU or trainable activations without smooth transition leads to a wide deviation of the final accuracy results and training failures. In contrast, using smooth transition improves the robustness of the training process, which succeeded for all seeds.

As described in Section \ref{section:ST}, the smooth transition starts at epoch $3$ and progresses over $10$ epochs. This explains the immediate accuracy drop at epoch $3$ for the left and middle graphs in Figure \ref{fig:robustness_squeezenet}. The rightmost graphs show the training process with a smooth transition, where the accuracy drop is delayed to the last transition epoch (around epoch $13$) or stretched during epochs $3-13$. One explanation for the accuracy drop in the rightmost graph is that the graphs of the trainable activations lay below the ReLU graph, see Figure \ref{fig:trainable_activations_graphs}. Therefore, when we set the activation to $\lambda_e \times trainable + (1-\lambda_e) \times ReLU$, the ReLU term is larger than the trainable term. Once the transition ends, we remove the ReLU term, which causes the model to respond with an accuracy drop. However, as we see in the graphs, the transition period helps stabilize the final accuracy. We also evaluated a smoother version of the transition, where we split the last $2$ epochs into $10$ smaller epochs with an increment of 0.02 for $\lambda_e$; but we did not observe any improvement.

\section{Conclusions} \label{sec:conc}
We introduced a new methodology for training HE-friendly models that replaces the ReLU activation functions with a trainable quadratic approximation. Our approach uses techniques such as polynomial activation functions with trainable coefficients, gradual replacement of activation layers during training, and \gls{KD}. In addition, our methodology can be automated and thus simplifies the way data scientists generate HE-friendly networks. Moreover, it allows them to avoid many struggles in achieving models with relatively good accuracy. We stress again that the entire training phase is done by the data owners on unencrypted data. Only when the model is ready the data owner encrypts it and uploads it to the cloud.

We tested our methodology on chest CT and \gls{CXR} image datasets using the AlexNet architecture, and showed that the performance of our trained model is only $0.32-5.3\%$ less accurate than the reference model, making it at least $15\%$ better than all previously suggested \gls{HE}-friendly training methods. We achieved similar results for SqueezeNet, which is a deeper network with 40 layers. Finally, we note that the authors of \cite{helayers} used our methodology to train AlexNet. Subsequently, they used their HELayers framework to demonstrate running it {\em over encrypted data} in less than five minutes without any degradation in accuracy. 

Secure computations using \gls{HE} is a rapidly growing domain and there are already several \gls{HE} frameworks that enable private \gls{DNN} computations on untrusted systems such as \cite{helayers, CHET2019}. However, they can provide accurate and usable solutions only in the presence of accurate \gls{HE}-friendly models. This puts our study on the critical path of deploying non-interactive HE-based solutions. In future work, we plan to further evaluate our methodology on more complicated models and domains. In addition, in the light of the results from SqueezeNet, we would like to design an  approach for replacing max-pooling that will offer less accuracy degradation. 

\bibliography{paper}
\bibliographystyle{splncs04.bst}

\appendix
\section{AlexNet Network Architecture} \label{appendix:a}

Our AlexNet architecture is the original implementation of PyTorch, where we added a batch normalization layer after every activation layer as our baseline. Here, all convolution layers use padding='same'. Note that the original model implemented by PyTorch has a \textit{Global Average Pooling} layer with an output size that is $6 \times 6$. For input images of size $224 \times 224$, it is equivalent to the identity function so we ignored it in our experiments.   

    

\paragraph{Our variant of AlexNet}

\begin{enumerate}
    \item Conv2d(3, 64, kernel=$11 \times 11$, stride=4, trainable\_polynomial)
    \item AvgPool2d($3 \times 3$, stride=2)
    \item BatchNorm2d(64)
    \item Conv2d(64, 192, kernel=$5 \times 5$, stride=1, trainable\_polynomial)
    \item AvgPool2d(kernel=$3 \times 3$, stride=2)
    \item BatchNorm2d(192)
    \item Conv2d(192, 384, kernel=$3 \times 3$, stride=1, trainable\_polynomial)
    \item Conv2d(384, 256, kernel=$3 \times 3$, stride=1, trainable\_polynomial)
    \item Conv2d(256, 256, kernel=$3 \times 3$, stride=1, trainable\_polynomial)
    \item AvgPool2d(kernel=$3 \times 3$, stride=2)
    \item BatchNorm2d(256)
    \item  Dropout(p=0.2)
    \item Flatten()
    \item FC(in=256, out=4096, activation=trainable\_polynomial)
    \item  Dropout(p=0.2)
    \item FC(in=4096, out=4096, activation=trainable\_polynomial)
    \item FC(in=4096, out=3)
\end{enumerate}
Total number of layers, including the activation layers and ignoring the dropout layers: 21.

\section{Model Hyperparameters} \label{appendix:b} 

We evaluated each experiment with 5 different seeds: 111, 222, 333, 444, 555. During the training process, we loaded images in mini-batches of size 32, and optimized the loss function using Adam optimizer. The number of epochs differs for each task, as does the learning rate, which was usually 3e-5 or 3e-4. 

The activation replacement started at epoch 3, and in case of smooth transition it was gradually replaced for 10 epochs. We replaced all ReLU activations in parallel. We found that it is better to initialize the coefficients with values similar to the form of ReLU, and that when the coefficents are scaled by a predefined number $(s_1,s_2)$, the network converges better. Therefore, the coefficients of each trainable activation were initialized as $s_1\times0.0X^2 + s_2\times 1.1x$ or $s_1\times 0.0X^2 + s_2\times 1.1x$, where $(s_1, s_2)$ are set to either $(0.1,0.1)$ or $(0.01,0.1)$. 

For the distillation process described in Section \ref{subsec:distil}, we set the temperature value to 10, and the $\alpha$ parameter was set to 0.1.

\end{document}